\documentclass[namedreferences]{kluwer}    
\newdisplay{guess}{Conjecture}
\usepackage{graphicx}
\begin{document}

\newcommand{\apj}{ApJ}
\newcommand{\apjl}{ApJL}
\newcommand{\apjs}{ApJSS}
\newcommand{\aap}{A&Ap}
\newcommand{\aj}{AJ}
\newcommand{\araa}{ARAA}

\begin{article}
\begin{opening}         
\title{Probing the Physics of Interacting Galaxies}

\author{Harriet \surname{Cullen}} 
\author{Paul \surname{Alexander}}
\institute{Astrophysics Group, Cambridge University, UK}
\author{Marcel \surname{Clemens}}
\institute{Dipartimento di Astronomia, Universita degli Studi di Padova, Italy}

\runningauthor{Harriet Cullen}
\runningtitle{Probing the Physics of Interacting Galaxies}


\begin{abstract}
The morphological and velocity structures in the gaseous ({\sc Hi} and CO) and stellar components of two
interacting systems are examined.  Both Arp 140 and Arp 104 reveal extended  tidal tails in the {\sc Hi}. 
The H$\alpha$ and FIR fluxes of Arp 140 yield similar SFR of $\sim$ 0.8 $ \mathrm{M}_{\odot}\,
\mathrm{yr}^{-1}$.  In contrast the H$\alpha$ flux of Arp 104 yields a SFR of $\sim$ 0.05 $
\mathrm{M}_{\odot}\, \mathrm{yr}^{-1}$, $\sim$ 20 times smaller than that obtained from the FIR flux.
Spectra were used to examine the changing velocity of atomic and molecular gas in NGC 5218 (Arp 104). The
atomic and molecular gas were found to be dynamically similar with comparable velocities and velocity
widths across the galaxy; consistent with the two phases responding similarly to the interaction, or
enhanced {\sc Hi} to CO conversion in the centre of the galaxy.
\end{abstract}

\end{opening}           

\section{Introduction}

A sample of galaxies consisting of one late and one early-type
galaxy is used to examine the effects of galaxy interaction on 
the ISM.  The sample has been selected to represent a range of
stages of interaction.  Selection of systems in which only one galaxy
is gas rich simplifies the modelling and interpretation. 
A combined observational modelling strategy is used, using morphological
and velocity structures observed in gaseous ({\sc Hi} and CO) and stellar components
to constrain the dynamics of these systems.  Here we present preliminary results for two systems; Arp 140
(NGC 274/275) and Arp 104 (NGC 5216/5218).
\section{Arp 140}  
                 
Arp 140 consists of two interacting galaxies; NGC 274 (SAB(r)0$-$ pec) and NGC 275 (SB(rs)cd pec).  At an
assumed distance of 26.9 Mpc, the projected major axis diameter corresponding to $D_{25}$ is 11.7 kpc
(1.5$'$) for both NGC 274 and NGC 275 and the projected separation of their centres is approximately 6.6
kpc (50$''$).  

We detect a large-scale distribution of {\sc Hi} around Arp 140 (Figure 1a).  The total measured {\sc Hi}
flux is 30.5 $\ \mbox{Jy\,km\,s}^{-1}$, corresponding to a mass of $5.1 \times 10^{9}$ M$ _{\odot}$.  An
extended {\sc Hi} tail reaches approximately $\sim 4'$ south of the optical galaxies and  $\sim 2.5'$
north, inconsistent with a direct collision between the two galaxies.  Over the area of 
optical emission of NGC 275, the line of sight velocity increases relatively uniformly from $\sim
1700\;\rm\mbox{km\,s}^{-1}$  on the south eastern side to $\sim 1875\;\rm\mbox{km\,s}^{-1}$ in the north
west, consistent with gas dynamically bound to the galaxy.  

\inlinecite{A140data} observed an integrated global H$\alpha$ luminosity of 9.54 $\times
10^{40}\,\mathrm{erg\,s}^{-1}$ which corresponds to a SFR of $\sim 0.75\, \mathrm{M}_{\odot}\,
\mathrm{yr}^{-1}$ ($SFR(\mathrm{H}\alpha)$/$\mathrm{M}_{\odot}$ \,$\mathrm{yr}^{-1} = 7.9 \times
10^{-42}L_{\mathrm{H}\alpha}$/$\mathrm{erg\,s}^{-1}$; \opencite{SFRhalpha}). At this star formation rate
the timescale for consumption of all gas associated with the galaxy is $\sim$ 7 Gyr.

A calculation of the star formation rate based on the far-infrared continuum yields a similar value of
$\sim 0.8\mathrm{M}_{\odot}\,\mathrm{yr}^{-1}$, ($\mathrm{SFR}(\mathrm{FIR})$/$\mathrm{M}_{\odot}$
\,$\mathrm{yr}^{-1} = 4.5 \times 10^{-44}L_{\mathrm{FIR}}$/$\mathrm{erg\,s}^{-1}$;  \opencite {SFRfir})
where the FIR luminosity is given by $L_{\mathrm{FIR}} \sim 1.7 \times\,L_{60}$ \cite{L60}.

\begin{figure}[hbtp]
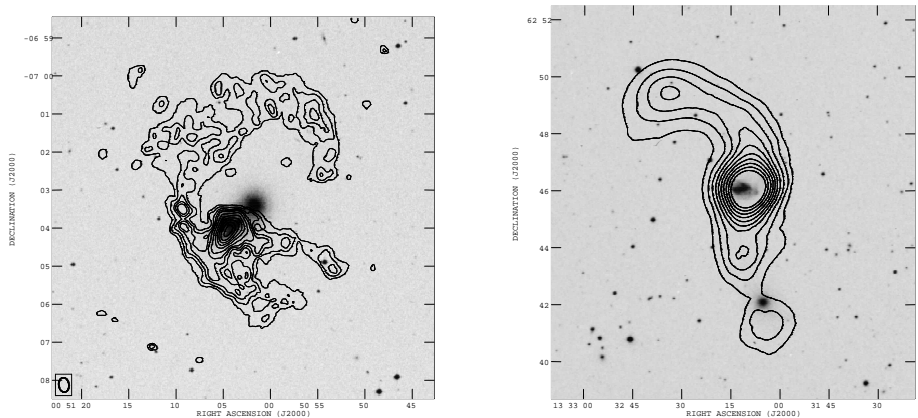

\centerline{\raisebox{5.5cm}{\includegraphics[width=5.5cm, angle=270]{cullen.fig1.eps}} \qquad
       {\includegraphics[width=5.5cm, angle=0]{cullen.fig2.eps}}}
\caption{(a), left, contours of $\Sigma_{{\sc Hi}}$ VLA C array data ($\theta_{\mathrm{FWHM}} = 23.01''
\times  15.96''$) for Arp 140.  Contour levels correspond to (0.03, 0.09, 0.15, 0.21, 0.27, 0.4, 0.6, 0.8,
1.0, 1.2, 1.4, 1.6) $\ \mbox{Jy beam}^{-1}\times \mbox{km\,s}^{-1}$.
(b), right, contours of $\Sigma_{{\sc HI}}$ VLA D array data ($\theta_{\mathrm{FWHM}}$ = 87.79$''$ $\times$
82.87 $''$) for Arp 104.  Contour levels correspond to (0.1, 0.2, 0.3, 0.4, 0.5, 0.6, 0.7, 0.8, 0.9, 1) $\
\mbox{Jy\ beam}^{-1}\times \mbox{km\,s}^{-1}$}
\end{figure}

\section{Arp 104}

Arp 104 consists of two interacting galaxies; NGC 5216 E0 pec (angular size of NGC 5216 is
$2.5'\times1.5'$), and NGC 5218 SB(s)b? pec (angular size $1.8'\times 1.3'$). At an assumed distance of
44.3 Mpc, the projected major axis diameter corresponding to $D_{25}$ for NGC 5216 is 32.2 kpc (2.5$'$) and
23.2 kpc for NGC 5218 (1.8$'$) and the projected separation of their centres is 52.9 kpc.

The measured {\sc Hi} flux is 17.1 $\mbox{Jy\,km\,s}^{-1}$, corresponding to a mass of $7.8\times 10^{9}$
M$_{\odot}$  The northern spiral galaxy appears disrupted with possible bar formation.  It has a tidal
plume extending to the north and a bridge to the south.  The southern galaxy is an elliptical with a small
tail in the south west.  These optical features are all reproduced in the low resolution {\sc Hi} map of
the Arp 104 system (Figure 1b).  A large tidal plume extends to the north east of NGC 5218 approximately 2
arcmin in length (26 kpc). Similarly, a bridge extends from the spiral to elliptical component, with an
increase in column density of the {\sc Hi} centred on the small tidal tail extending to the south west of
the elliptical.

\inlinecite{schom} have conducted a multicolour photometric study of the tidal features of interacting
galaxies which included Arp 104.  They found the colours of all the tidal features to be bluer than those
of the central regions of either galaxy, but similar to the colours of the outer disk of the spiral
component.  Given the smooth appearance of the tidal features they propose a scenario in which material is
being removed from the disk of NGC 5218.

\inlinecite{A104data} observed an H$\alpha$ luminosity of 6.0 $\times 10^{39}\,\mathrm{erg\,s}^{-1}$ giving
a SFR of $\sim 0.05\, \mathrm{M}_{\odot}\, \mathrm{yr}^{-1}$ (Kennicutt et al 1994) and a timescale for
consumption of all gas associated with the galaxy of $\sim$ 170 Gyr. However, a calculation of the star
formation rate based on the far infrared continuum yields a much larger star formation rate of $\sim 1.2\,
\mathrm{M}_{\odot}\, \mathrm{yr}^{-1}$ which would reduce the timescale for gas consumption to $\sim$ 7
Gyr. The difference in star formation rates can be reconciled if there is  $\sim$ 3.5 magnitude of
extinction towards the star forming regions.

\subsection{Velocity Structure of {\sc Hi} and $^{12}$CO$(J=2\to1)$ in Arp 104 }
{\sc Hi} data from the VLA and $^{12}$CO$(J=2\to1)$ data from the JCMT have been used to determine spectra
of the two gas phases at 10 arcsec intervals over a 50$''\times$ 90$''$ area, centred on NGC 5218, the
spiral component of the interacting pair.  The data are fully sampled with matched spatial and velocity
resolution.

We use these data to examine the changing velocity of the atomic and molecular gas through the galaxy: 
velocities of each component are plotted against radial distance from the galaxy centre in Figure 2. 
Comparison of the velocity fields is complicated by two factors, (1) there is evidence that the {\sc Hi}
becomes optically thick with flat spectra making location of a peak velocity difficult and (2) much of the 
{\sc Hi} appears in absorption over the region where there is strong CO emission. Considering only those
locations in which reliable velocities for both phases are obtained there is good evidence that the atomic
and molecular gas are dynamically similar; the velocities and velocity widths of the two gas phases being
comparable across the galaxy.  This is consistant with the atomic and molecular gas phases having responded
similarly to the interaction, or with an enhanced {\sc Hi} to CO conversion at the centre of the galaxy.

\begin{figure}[hbtp]
\centering\includegraphics[width=4cm, angle=270]{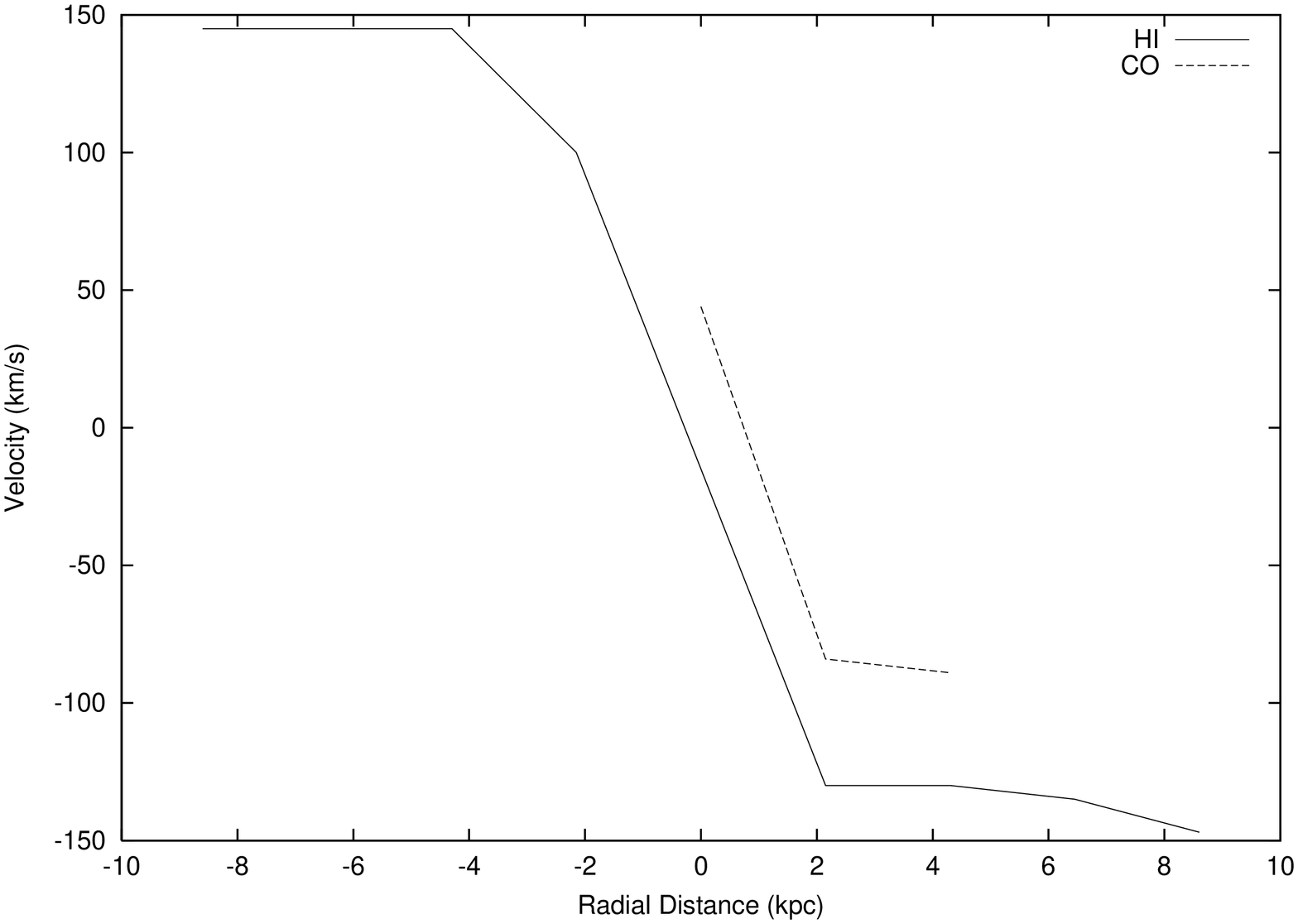}
\centering\includegraphics[width=4cm, angle=270]{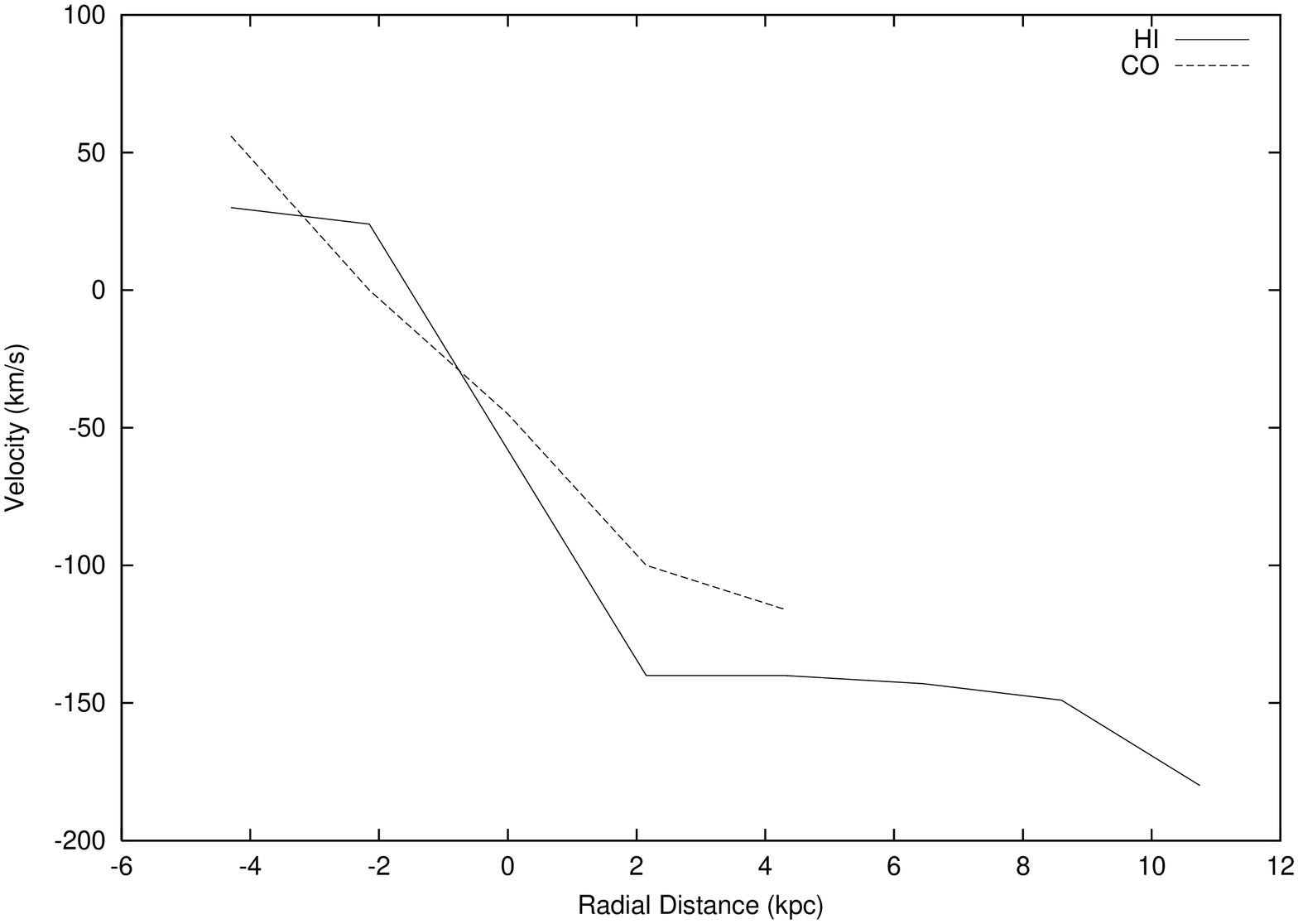}
\caption{Plot showing the line of site velocity of the atomic and molecular phases of NGC 5218 against the
distance in kiloparsecs from the centre of the optical galaxy for two cuts through the galaxy}
\end{figure}

\bibliographystyle{klunamed}
\bibliography{cullen}

\acknowledgements
The VLA is operated by the NRAO for AUI. The
DPSS was funded by the National Geographic society and produced by the Space Telescope Science Institute
from plates taken with the Oschin
Schmidt Telescope. This is operated jointly by the California Institute of Technology and the Palomar
Observatory. The JCMT is operated by The JAC on behalf of the PPARC on behalf of the UK, the Netherlands
Organisation for Scientific Research, and the National Research Council of Canada.


\theendnotes

\end{article}
\end{document}